\documentclass[pre,aps,twocolumn,showpacs,floatfix]{revtex4}
\usepackage[utf8]{inputenc}
\usepackage{amsmath}
\usepackage{amsfonts}
\usepackage{amssymb}
\usepackage{graphicx}
\usepackage{subfigure}
\usepackage{ulem}
\usepackage{hyperref}
\usepackage{color}
\usepackage{tikz}
\usetikzlibrary{decorations.pathmorphing,patterns,arrows,shapes.misc}
\usepackage{tikz-3dplot}
\usepackage[nomessages]{fp}
\usepackage[first=0, last=360]{lcg}
\usepackage{ifthen}
\usepackage{csvsimple}
\usepackage{comment}

\tikzset{cross/.style={cross out, draw=black, minimum size=3*(#1-\pgflinewidth), inner sep=0pt, outer sep=0pt},
	cross/.default={6}}

\begin{document}

\title{Radial Restricted Solid-on-Solid and Etching Interface Growth Models}
\author{Sidiney G. Alves}
\email{sidiney@ufsj.edu.br}
\affiliation{Departamento de F\'isica e Matem\'atica, Universidade Federal de S\~ao Jo\~ao Del-Rei \\
	36420-000, Ouro Branco, MG, Brazil}

\begin{abstract}
In this work an approach to generate radial interfaces is presented. A radial network recursively obtained is used to implement discrete model rules designed originally for the investigation in flat substrates. In order to test the proposed scheme, we have used the restricted solid-on-solid and etching models. The results indicate the KPZ conjecture is fully verified. Besides, a very good agreement between the interface radius fluctuation distribution and the GUE one was observed. The evolution of the radius agrees very well with the generalized conjecture, and the two-point correlation function exhibits a very good agreement with the covariance of Airy$_2$ process. So, this approach can be used to investigate radial interfaces evolution for others universality classes.
\end{abstract} 

\maketitle

\section{Introduction}

Fluctuations are inherent to far from equilibrium phenomena, and are present in a broad range of growing systems in 
nature, from thin film deposition to biological growth. These fluctuations are characterized
by self-similarity and universality that emerge from distinct dynamical processes of their formation \cite{barabasi,meakin}.
A large number of  non-equilibrium fluctuations phenomena are associated
to the universality class of the equation proposed by Kardar, Parisi, and Zhang (KPZ) \cite{KPZ}
\begin{eqnarray}
\frac{\partial h(x,t)}{\partial t} = \nu \nabla^2 h(x,t) + \lambda \left( \nabla h(x,t) \right)^2 + \eta(x,t)
\end{eqnarray}
where $h(x,t)$ represents the interface height at a position $x$ in time $t$, the first term in the right hand 
corresponds to surface tension, the non-linear second term represents a local lateral growth in the
normal direction along the surface and the last one is a white noise with $\langle \eta(x,t)\rangle = 0$ and 
$\langle \eta(x,t) \eta(x',t')\rangle = D \delta(x-x')\delta(t-t')$.

In the last two decades, we witnessed an increase in the interest to the KPZ universality class  due to the exact results
obtained for $1+1$ dimensions \cite{johansson,PraSpo1}. The central point of these exact results is that the height
fluctuations of interfaces belonging to the KPZ universality class can be described by Tracy-Widom distributions. The fluctuation probability distribution function depends on the substrate geometry or initial condition dividing the KPZ universality in different sub-classes. The height evolution of a single site is given by  
\begin{equation}\label{eq:height}
 h(t) = v_\infty t + (\Gamma t)^\beta \chi
\end{equation}
where $v_\infty$ and $\Gamma$ are system depending parameters, $\beta$ is the
universal growth  exponent and $\chi$ is a random variable that describes the fluctuations. The probability distribution function
of this random variable splits the KPZ universality in two sub-classes depending on the substrate geometry. In flat substrates,
the random variable is associated to the Gaussian orthogonal ensemble distribution, while, in curved substrates, the random
variable follows the Gaussian unitary ensemble (GUE) distribution. This conjecture was verified in several analytical (including exact solution) 
\cite{SasaSpo1,SasaSpohnJsat,Amir,CalaDoussal,Imamura}, experimental \cite{TakeSano,TakeuchiSP,TakeSano2012,yunker}
and numerical \cite{SchehrEPL,Alves11,Oliveira12,Alves13,bd_box2d} works for both geometries. The conjecture stated by the equation \ref{eq:height}
was extended to high dimensions. Initially in $2+1$ dimensions, with numerical approaches \cite{healyPRL,Oliveira12R,bd_box2d} and followed by experimental 
evidences \cite{healy2014,almeida2013,Almeida2015}. Later, it was extended to dimensions up to $6+1$ using numerical simulations of
discrete growth models \cite{RSOShigh,BDbox34}.

The conjecture in Eq. (\ref{eq:height}) was generalized to take into account finite time corrections in the height cumulants 
observed in various works using analytical ~\cite{SasaSpohnJsat,Amir,CalaDoussal,SasaSpo1,Ferrari},
experimental~\cite{TakeSano2012,TakeSano,TakeuchiSP} and numerical approaches~\cite{Alves11,Oliveira12,Oliveira12R,bd_box2d,Takeuchi12,Alves13}. The generalized version includes additional terms as follows
\begin{equation}
h(t) = v_\infty
t+s_\lambda(\Gamma t)^{1/3}\chi+\eta+\zeta t^{-1/3}+\ldots. 
\end{equation}
here $\eta$ and $\zeta$ are non-universal parameters. Both play an important role at finite-time analyses.

The numerical investigation of interface growth is traditionally based on discrete growth models. The main investigated
models belonging to the KPZ universality class are restricted solid-on-solid (RSOS)~\cite{rsos}, ballistic
deposition~\cite{vold}, single step~\cite{single_step}, etching~\cite{etching} and Eden~\cite{eden}.
In the particular case of a curved substrate, besides all the works mentioned above, the main studies 
are based on radial interface evolution of the Eden growth model~\cite{eden} and its variations~\cite{Alves13}.
On lattice simulations of the radial version exhibits anisotropy effects leading to
distorted interfaces with growth velocity varying along the chosen direction. The analysis
must be done using fixed directions~\cite{Alves13}. Even though the
off-lattice radial simulations in $1+1$ dimension are quite affordable, large scale simulation are expensive~\cite{BJP}.

In this work, a method to adapt growth model rules to generate radial interfaces is proposed. 
The rules of RSOS and etching models were used in a lattice built hierarchically. The obtained results for both models indicate a convergence of the universal parameter of KPZ universality class. The KPZ conjecture is fully verified with a very good agreement 
between the interface radius fluctuation distribution and the GUE one. Besides, the results obtained 
to the two-point correlation function exhibits a very good agreement with the covariance of the Airy$_2$ process.

The paper is organized as follows. In the next section, the hierarchical network is described, and the growth
rules details are presented. In Sec. \ref{sec:results}, we present and discuss the results. Section \ref{sec:conclusions} is devoted to the conclusions.

\section{Model}
\label{sec:model}
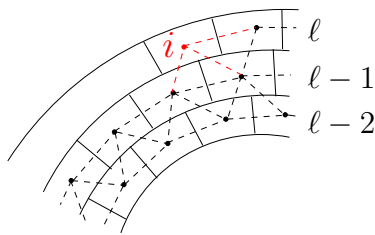
\begin{figure}[!b]
\begin{center}
\includegraphics[width=0.45\linewidth]{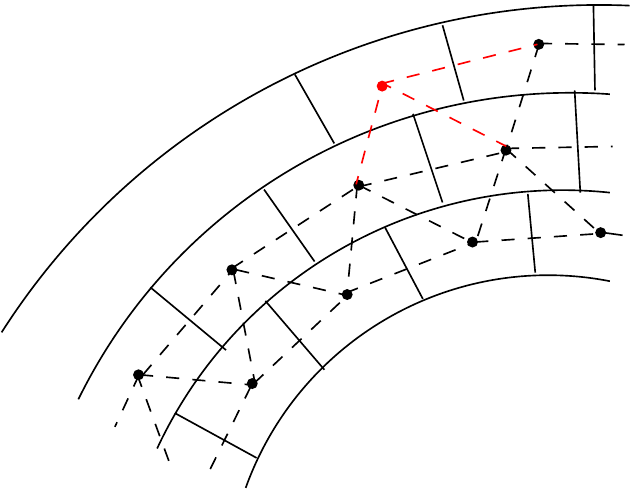}
\begin{picture}(0,0)
\put(-55,67){\color{red}\Large $i$}
\put(0,73){\large $\ell$}
\put(0,55){\large $\ell-1$}
\put(0,38){\large $\ell-2$}
\end{picture}
\end{center}
\caption{\label{fig:net_build}(Color online) Schematic illustration of the construction of the neighborhood in the radial network. The last added site and the bonds with the site of its layer and of the previous one is highlighted in red. Notice that the bond with the site of the layer and with that of previous layer is identified in the addiction.}
\end{figure}

The radial network was built considering a set of circular concentric layers enumerated as $\ell =0$, $1$, $\ldots$ A
layer $\ell$  has a radius $r_{\ell}$ and $N_\ell = \mbox{INT}(2\pi r_\ell)$ sites here the radius $r_\ell=r_0+\ell$ 
and $INT(x)$ is a function to return the value of $x$ truncated. The $N_\ell$ sites are distributed on the
layer $\ell$,  in a way that its cell has an arc length of unitary size.
The site locations are obtained recursively beginning with a random selection of the angular location $\phi_1$ of
the first one, the remaining sites of the layer are positioned
considering $\phi_i = \phi_{i-1}+\delta\phi$ with $i=2,3,\ldots N_\ell$  and 
$\delta\phi=\frac{2\pi r_\ell}{N_\ell}$. 
The neighborhood of the sites is obtained during the construction process as schematically 
illustrated  in Fig. \ref{fig:net_build}.
A new site $i$ at a layer $\ell$ will be a neighbor of the site $i-1$ at same layer and of those at the inner layer ($\ell - 1$).
In this case, we search by the sites of $\ell - 1$ layer that share an edge with the cell of site $i$. 
So, in general, each new site has bonds with three neighbors previously added. Besides, the neighborhood of 
site $i-1$ and of those on layer $\ell-1$ are updated to store the site $i$ as neighbor.
The procedure is repeated to obtain the interested number of layer.

\begin{figure}[!t]
	\includegraphics[width=0.75\linewidth]{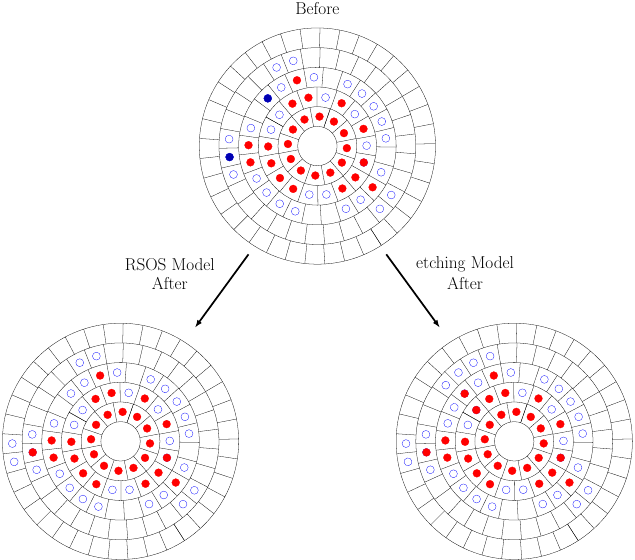}
	\caption{(Color on-line) \label{fig:growth_rules}Schematic illustration of the growth rules of the RSOS and etching models. The top panel show two chosen sites (blue full circles). Bottom panels show the new configurations using the RSOS (left) and the etching rules (right).}
\end{figure}

The interface growth rules for RSOS and etching models are implemented considering, as initial condition, all sites of the first layer occupied.  The top panel in Fig.~\ref{fig:growth_rules} shows an illustration of an interface after few steps. The deposited and peripheral sites are highlighted by red full and blue open circles, respectively. New particles will be deposited on the peripheral sites. 
The RSOS rules are implemented considering that a new incoming particle will occupy a peripheral site at random. The incorporation of the new particle to the interface must take into account the previous layer. If all neighboring sites of this layer are occupied the deposition is accepted. Otherwise, if the site $i$ has at least one empty neighbor at the previous layer, the deposition is refused. These implementation rules are equivalent to the flat deposition considering the restriction parameter $m=1$.
The etching growth rules are implemented as follows. A deposition site $i$ belonging to the peripheral is chosen. All empty neighbors in previous layers are filled. 
The update rules for the two models are schematically illustrated in Fig.~\ref{fig:growth_rules}. The upper configuration shows two peripheral sites chosen to deposition (they are highlighted by blue filled circles). The bottom configurations show the interface after the update using RSOS (left) and etching (right) rules. 
Since we randomly pick up a peripheral site from a constantly updated list containing $N_p$ sites, the time is updated as $t=t+dt$ at each attempt with $dt=1/N_p$. This strategy to update the time is based on that used in the Eden growth model where only the peripheral sites may divide \cite{Alves11}.
The simulations were carried out on networks considering the first radius layer $r_0 = 10$ and the simulations run until the aggregates reach a radius $r=10^4$. Averages are take from up to $N = 10^4$ independent samples. 

\section{Results and discussion}
\label{sec:results}

The investigation of the radial growth was done considering the time evolution of the interface radius fluctuations. 
The cells belonging to the interface at time $t$ are defined by the set of those adjacent to the peripheral one. As
mentioned previously, the radius of an interface cell is defined by its layer, {\it i.e.}, a cell in the layer $\ell$ has a radius
$r_\ell = r_0+\ell$. 

\begin{figure}[!b]
	\subfigure[\label{fig:rsos1}]{\includegraphics[width=0.9\linewidth]{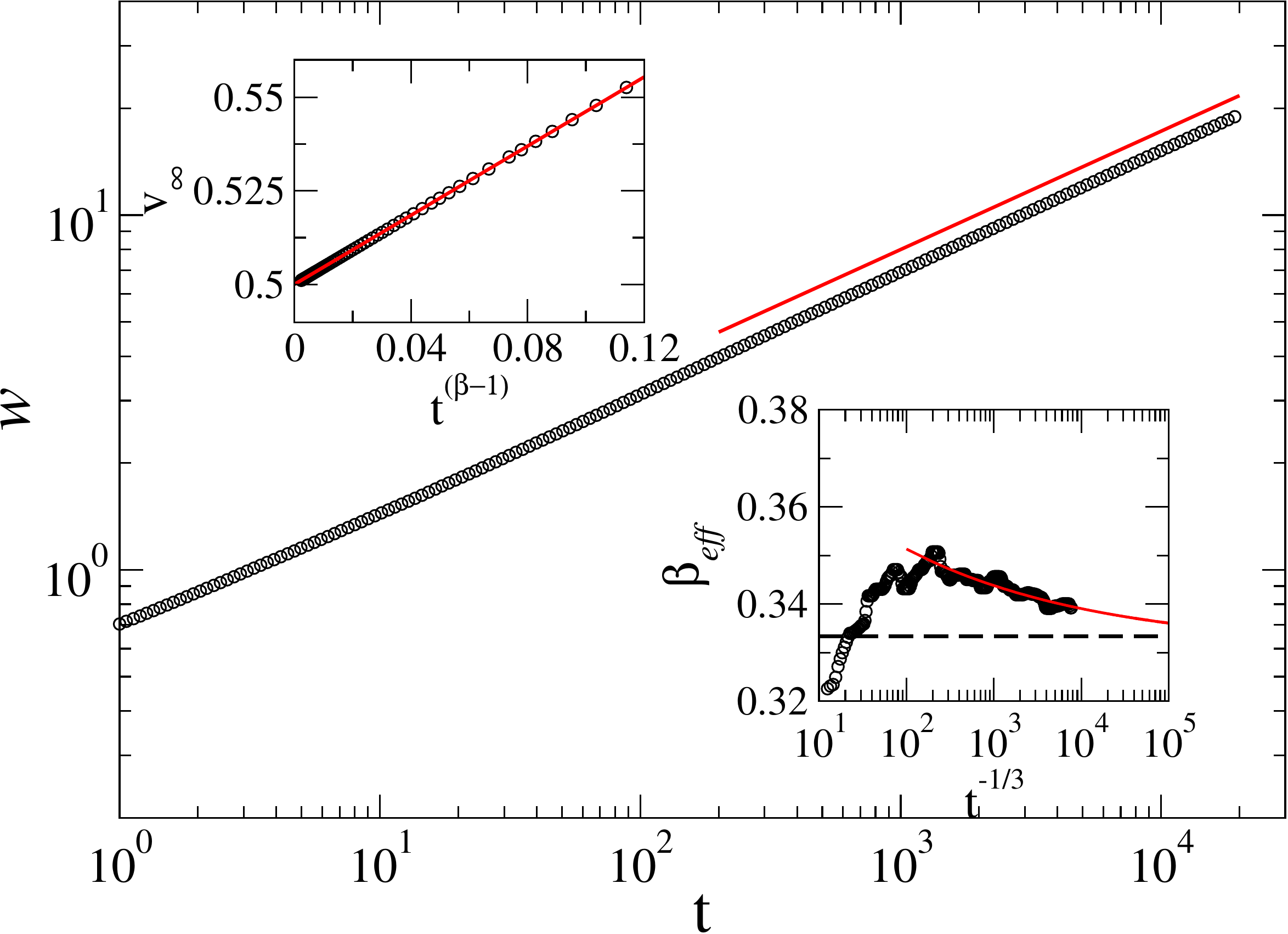}}
	\subfigure[\label{fig:etch1}]{\includegraphics[width=0.9\linewidth]{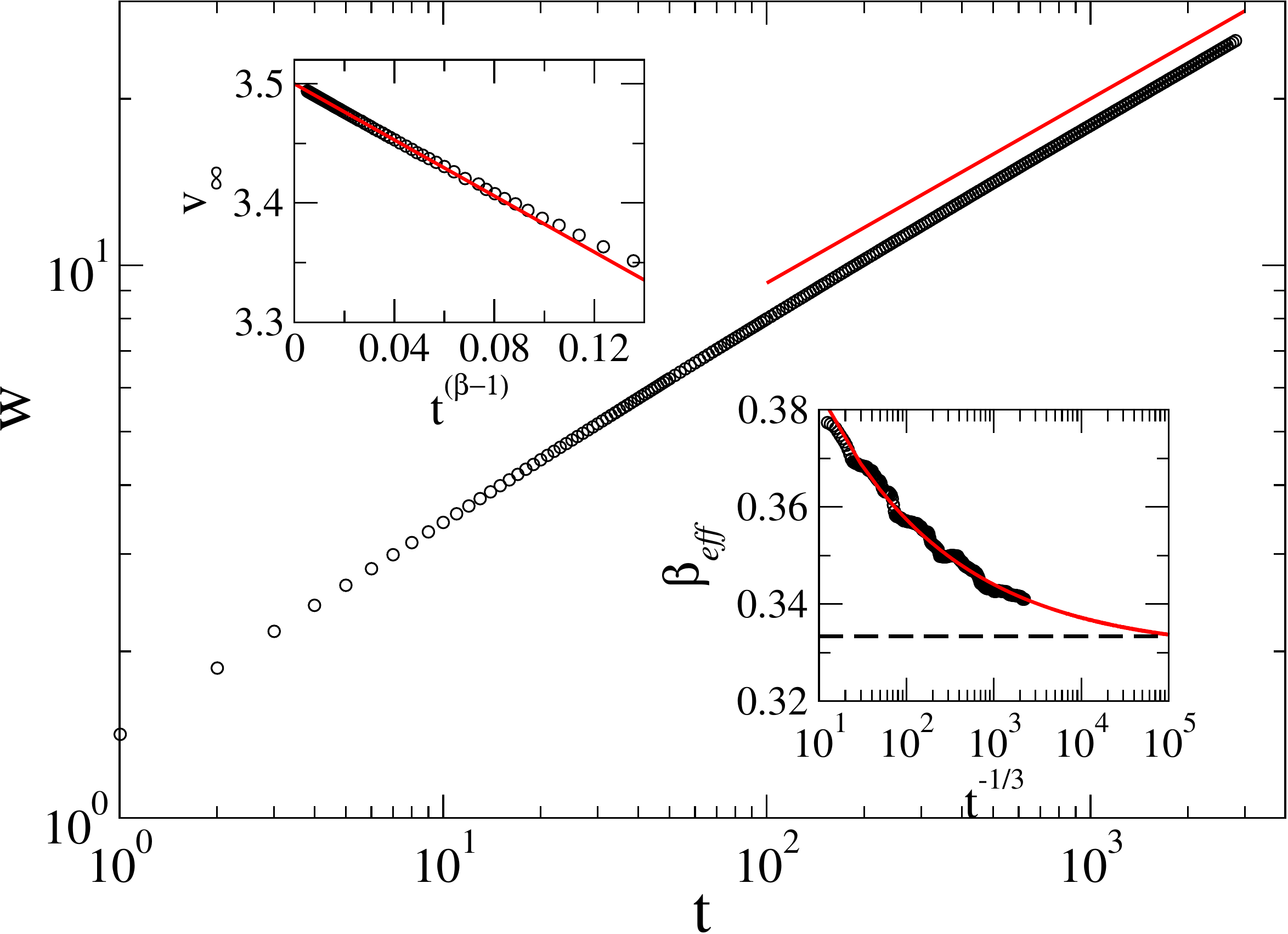}}
	\caption{\label{fig:w_vinf_beta}(Color online) Evolution of the second cumulant of the radius for the RSOS (a) and etching (b) models. Insets show the time derivative of the first (top insets) and of the second (bottom insets) cumulant for the two models.}
\end{figure}

At short times, the height fluctuation exhibits a behavior with Gaussian distribution as reported previously \cite{Prolhac}
(results not show).
The asymptotic value of the interface growth velocity is obtained considering the time derivative of the first moment. From
Eq. (\ref{eq:height}), it is given by
\begin{equation}\label{eq:dh}
\partial_t \langle h \rangle = v_\infty + s_\lambda \Gamma^\beta \beta t^{\beta - 1} \langle\chi\rangle.
\end{equation}
So, as shown in the upper insets of Fig. \ref{fig:rsos1} and \ref{fig:etch1}, the value of $v_\infty$ is obtained using a linear fit in the plot of $\partial_t \langle h \rangle $ against $t^{\beta-1}$ here $\beta=1/3$ was used. The asymptotic values obtained for both, RSOS and etching growth model, are presented in table \ref{tab:pars}. 
Notice that, considering Eq. \ref{eq:height}, the radius second cumulant is given by
\begin{equation}\label{eq:roughness}
\langle h^2\rangle_c = (\Gamma t)^{2\beta} \langle\chi^2\rangle_c
\end{equation} 
The results in main panels of Fig. \ref{fig:rsos1} and \ref{fig:etch1} show a clear power law $w=(\langle h^2\rangle_c)^{1/2}$ vs time for both RSOS and etching models. Bottom insets of these figures show $\beta_{eff}$ as a function of $t^{-1/3}$. Here, $\beta_{eff}=d(ln(w))/d(ln(t))$ gives the local derivative of $ln ~ w$ versus $ln ~ t$. For the RSOS model, a non-monotonic behavior is observed. However, in both models the growth exponent asymptotically converges to values very close to $1/3$. 
The obtained values are presented in table \ref{tab:pars}. In addiction, the higher order cumulants were used to obtains the skewness ($S$) and kurtosis ($K$)  associated to the radius fluctuations, defined as
\begin{eqnarray}
S & = & \frac{\langle{h^3}_c\rangle}{\langle{h^2}_c\rangle^{1.5}} ~\\
K & = & \frac{\langle{h^4}_c\rangle}{\langle{h^2}_c\rangle^2}
\end{eqnarray}
The plot of $S$ and $K$ against time exhibits a finite time correction in the convergence to the asymptotic value (results not shown) consistent with a power law convergence with exponent $2/3$. The estimated values are in very good agreement with those associated to the KPZ class (see Table~\ref{tab:pars}).

\begin{table}[!t]
	\scalebox{0.85}{
		\begin{tabular}{c||c|c|c|c|c|c|c}
			~ &  $v_\infty$ & $\beta$ & $\Gamma_1$ & $\Gamma_2$ & $\langle\eta \rangle$&  $S$ & $K$ \\ \hline\hline
			RSOS     & 0.50001(4) & 0.334(3) & 0.49(1) & 0.50(1) & 8.2(2)  & 0.22(1) & 0.095(2) \\ \hline
			etching  & 3.49997(3) & 0.331(2) & 8.2(2)  & 8.5(3)  & 15.4(3) & 0.21(1) & 0.085(5)
	\end{tabular}}
	\caption{\label{tab:pars}Nonuniversal and universal quantities for the dynamical regime of KPZ models.}
\end{table}

\begin{figure}[t]
	\includegraphics[width=0.9\linewidth]{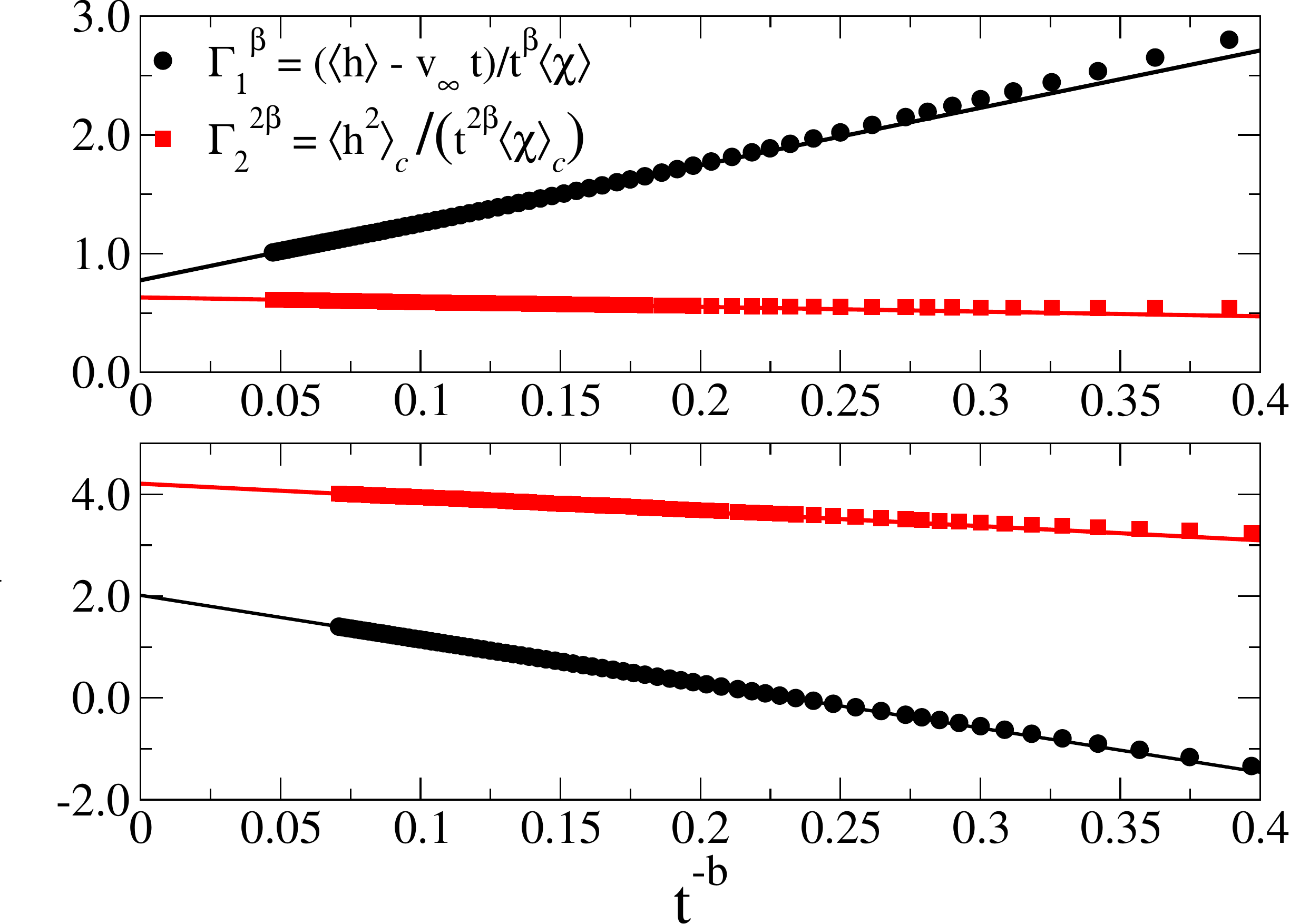}
	\caption{\label{fig:gammas}(Color online) The $\Gamma$ parameter obtained using the Eq. (\ref{eq:gamma1}) and (\ref{eq:gamma2}) for RSOS and etching models, top and bottom, respectively. Here, we set $\langle\chi\rangle = -1.771069$ and $\langle\chi^2\rangle_c = 0.812729$ and $b=\beta$ and $2 \beta$ for the $\Gamma_1$ and $\Gamma_2$, respectively.}
\end{figure}

A crucial step is the determination of the parameters $\lambda$ and $\Gamma$ to analyze the agreement of the radius fluctuations distribution with the GUE one. 
In the radial interface growth $\lambda = v_\infty$ and the $\Gamma$ parameter are derived from Eq. (\ref{eq:height}) and (\ref{eq:roughness}) and reads
\begin{eqnarray}
\label{eq:gamma1}\Gamma_1^\beta & = & \frac{\langle h\rangle - v_\infty t}{t^\beta \langle\chi\rangle} \\
\label{eq:gamma2}\Gamma_2^{2\beta} & = & \frac{\langle h^2\rangle_c}{t^{2\beta} \langle\chi^2\rangle_c},
\end{eqnarray}
regarding that $\langle\chi\rangle = -1.771069$ and $\langle\chi^2\rangle_c = 0.812729$ are the first and second cumulants of the GUE distribution. The values of $\Gamma_1$ and $\Gamma_2$ were obtained using a linear extrapolation in the plot against $t^{-\beta}$ and $t^{-2\beta}$, respectively, as shown in Fig.~\ref{fig:gammas} (solid lines). The estimated values are presented in table \ref{tab:pars}. Note that the values of $\Gamma_1$ and $\Gamma_2$ are close for both models, so in the next results we consider $\Gamma$ as a mean of these two values.

\begin{figure}[b]
	\includegraphics[width=0.9\linewidth]{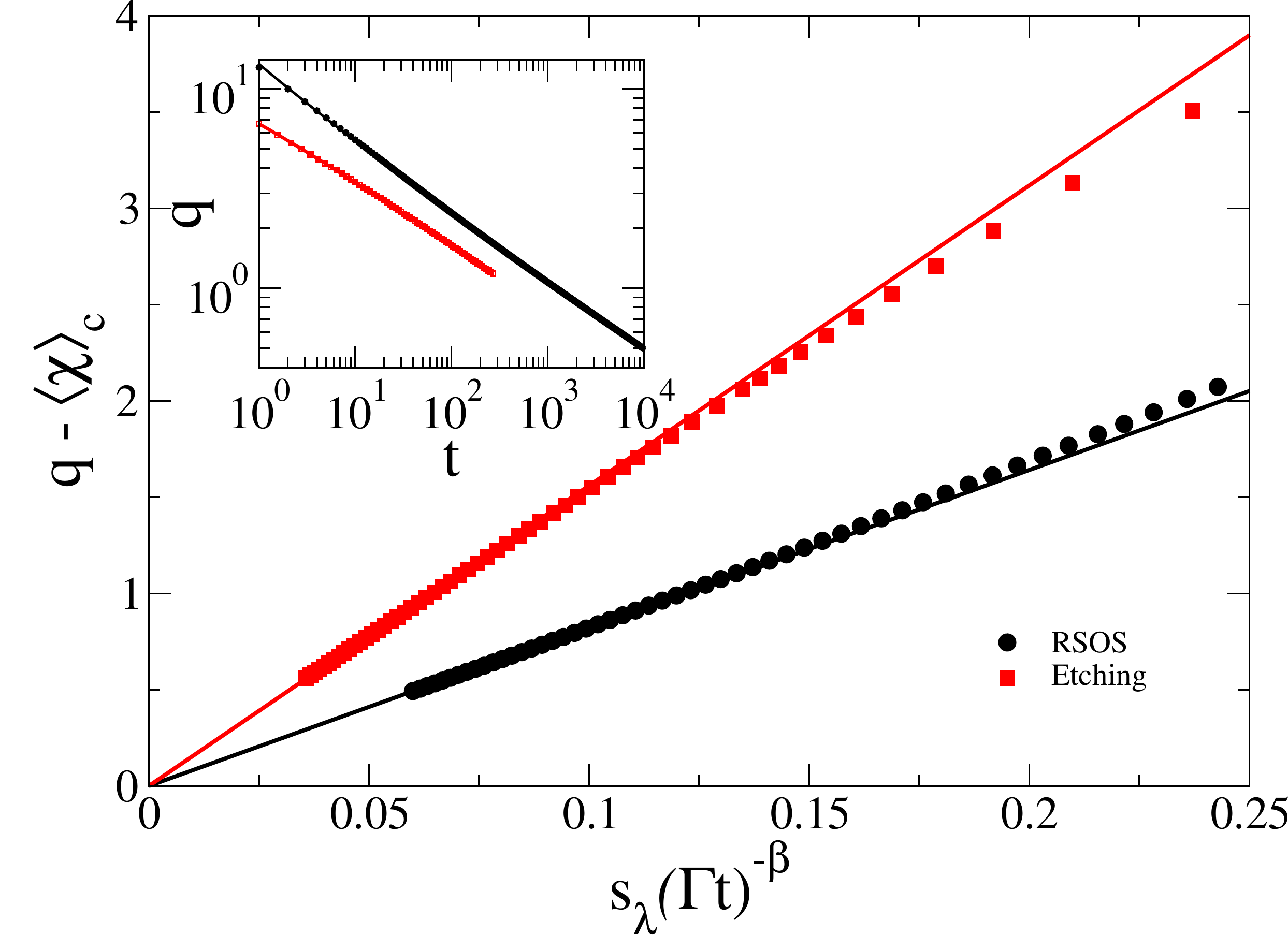}
	\caption{\label{fig:shifts}(Color online) Convergence of the random variable $q$ to $\langle\chi\rangle$ for the two models as shown in the legend. The solid lines are linear fit to the asymptotic regions.}
\end{figure}

Using the estimated parameters in the previous analysis and the Eq. (\ref{eq:height}) we can calculate the random variable given by
\begin{equation}
q = \frac{\langle h\rangle - v_\infty t }{s_\lambda(\Gamma t)^\beta}.
\end{equation}
The plot of $q -\langle\chi\rangle $ against $s_\lambda (\Gamma t)^{-\beta}$ is shown in Fig.~\ref{fig:shifts}. A clear linear behavior is observed asymptotically. Besides, when the evolution from early times is considered, as can be seen from inset in Fig.~\ref{fig:shifts}, a very good fit is obtained when a double power law $q - \langle\chi\rangle=a t^{-1/3} + b t^{-2/3}$ is used in the plot. This regression was considered previously in numerical simulations~\cite{Alves13}. 
So, the $q$ variable converges to $\langle\chi\rangle$ as
\begin{equation}
q =  \langle\chi\rangle + \frac{\langle\eta\rangle}{s_\lambda(\Gamma t)^\beta}+\frac{\langle\zeta\rangle}{s_\lambda\Gamma^\beta{t^{2\beta}}}
\end{equation}
as aforementioned, here the second term is associated to a shift in the $q$ variable in relation to the mean value of the GUE distribution and the last term to finite time corrections. In the light of the previous results, the behavior of the radius is in agreement with the generalized KPZ conjecture.

\begin{figure}[b]
	\subfigure[\label{fig:p_qprime_rsos}]{\includegraphics[width=0.9\linewidth]{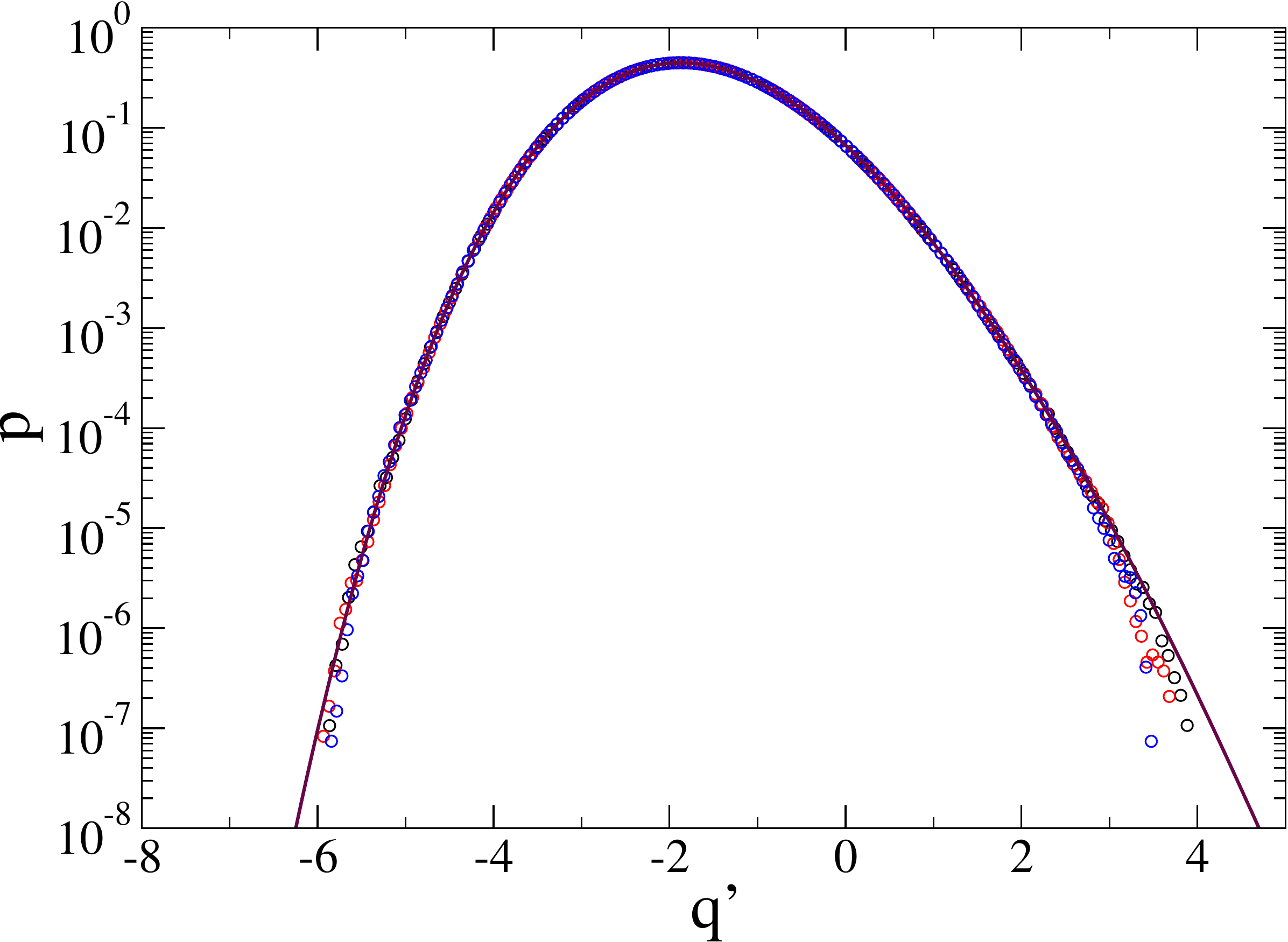}}
	
	\subfigure[\label{fig:p_qprime_etch}]{\includegraphics[width=0.9\linewidth]{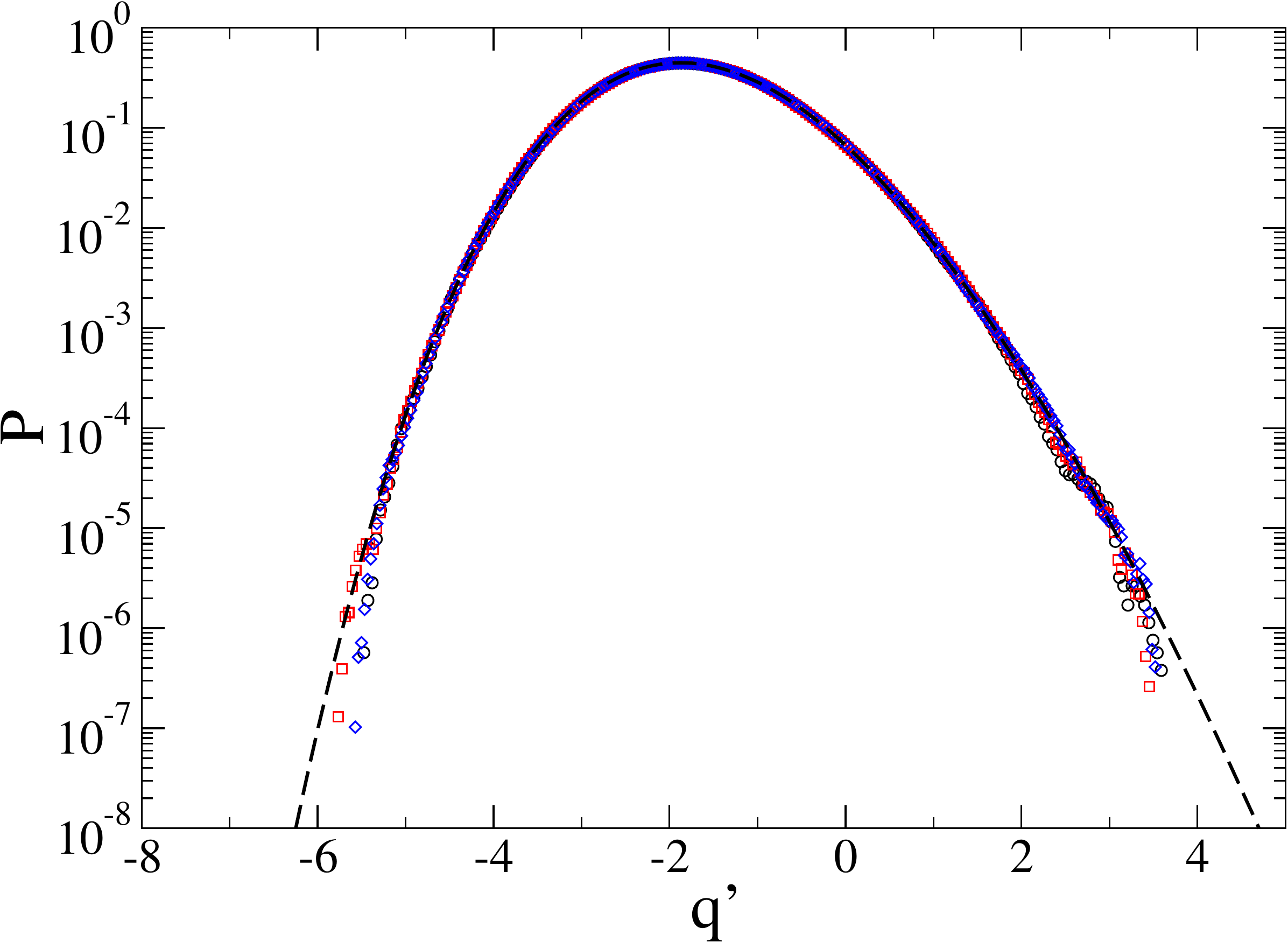}}
	\caption{\label{fig:pqprime}(Color online) Height distribution function at different times scaled according to the Eq.~(\ref{eq:qprime}). For the RSOS model we have used $t=5000$, 7500 and $10^4$ while for the etching model, $t=10^3$, 2000 and 3000.}
\end{figure}

To compare the obtained radius distribution function of RSOS and etching models with the TW-GUE one, the radius can be scaled as
\begin{equation}
\label{eq:qprime}q' = \frac{\langle h\rangle - v_\infty t -\langle\eta\rangle }{s_\lambda(\Gamma t)^\beta}.
\end{equation}
Figures~\ref{fig:p_qprime_rsos} and \ref{fig:p_qprime_etch} show the rescaled probability distribution for RSOS and etching models, respectively. An excellent agreement is observed for the two models.

\begin{figure}[!h]
	\includegraphics[width=0.9\linewidth]{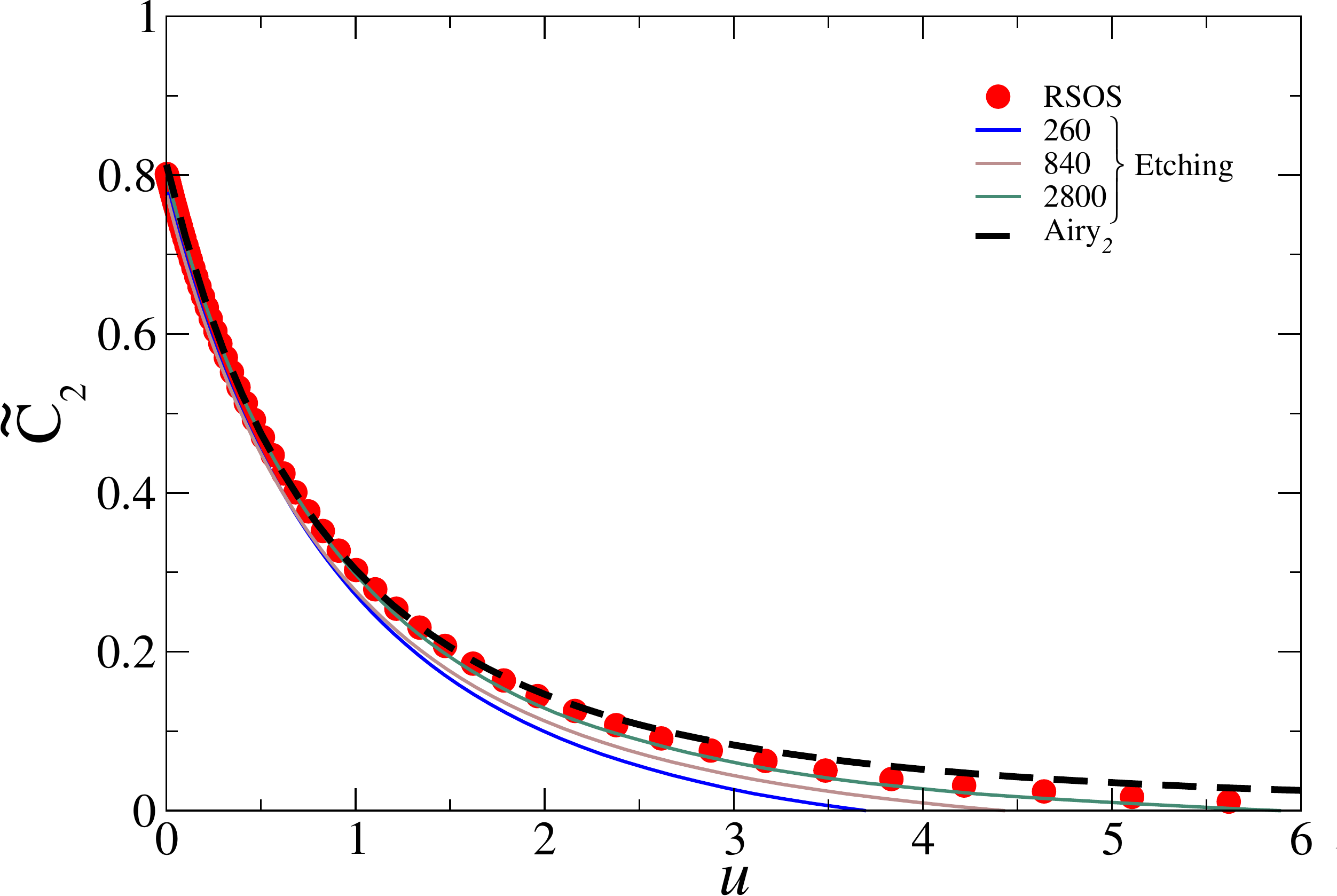}
	\caption{\label{fig:c2_airy2}(Color online) Rescaled two-point correlation function $\tilde{C}_2 = (\Gamma t)^{-2\beta}\times C_2$ against the rescaled length $u$. The dashed line represents the Airy$_2$ correlation function.}
\end{figure}

Finally, the two-point correlation function, given by
\begin{equation}
C_2(\varepsilon,t) = \langle r(x+\varepsilon, t) r(x,t) \rangle - \langle R(x,t)\rangle
\end{equation}
for radial growth models belonging to the KPZ universality exhibit the scaling $C_2(\varepsilon,t) \approx (\Gamma t)^{2\beta} g_2(u)$ with $u=(A\varepsilon/2)(\Gamma t)^{2\beta}$ and $g_2(u)$ is the covariance of the Airy$_2$ process~\cite{bornemann}. To verify the scaling for the RSOS and etching interfaces we plot in Fig. \ref{fig:c2_airy2} the two-point correlation function considering the rescale $\tilde{C}_2 = (\Gamma t)^{-2\beta}\times C_2$ against $u$. A remarkable agreement is observed for the RSOS model while the results for the etching are converging (the time is increased from bottom to top curves as indicated on the legend figure).

~

\section{Conclusions}
\label{sec:conclusions}

In this work, we present a strategy to investigate the discrete interface growth models in radial geometry.
The restricted solid-on-solid (RSOS) and etching growth model rules were adapted to generate radial interfaces. The fluctuations of the interface radius was investigated in $1+1$ dimensions. 
The radius moments, growth exponent and universal quantities presented an excellent agreement with those of the KPZ universality class. Besides, the generalized KPZ conjecture was fully verified with  the
radius distribution exhibiting a very good agreement with the Tracy-Widom of the Gaussian unitary ensemble, as conjectured to the KPZ universality class for curved surfaces. These finding were corroborated by the quantitative agreement between the RSOS and etching two-point correlation scaling and the Airy$_2$ process. Recently, quantitative predictions for the universal form of the two-time correlations in the infinite time limit of the KPZ equation were derived \cite{Nardis}. This work show the breaking of the ergodicity in radial systems time evolution. So, these universal properties will be investigated using the present strategy on a future work. Furthermore, the strategy present here can be used to investigate the interface growth of models belonging to others universality classes.
We are working in the implementation of the rule of conserved version of the RSOS and Das Sarma-Tamborenea models, that belong to the nonlinear molecular beam epitaxy universality class. A recent work show that these models present a dependency on flat (fixed-size) and expanding substrates \cite{Carrasco}. The results will appear elsewhere as soon as possible.

~\\

\section*{Acknowledgements}

The author thanks the discussions with Silvio C. Ferreira and for the critical reading of the manuscript by him and Marcelo M. Oliveira. The author also thanks F. Bornemann by kindly providing the covariance of the Airy$_2$ process. This work was partially supported by CNPq and FAPEMIG (Brazilian agencies).


\end{document}